\documentclass{statsoc}

\advance\hoffset by -1in
\advance\voffset by -1in

\usepackage{mathtools}
\usepackage{natbib}
\newcommand{\argmax}{{\rm argmax}}
\newcommand{\Perp}{\perp \! \! \! \perp}
\newcommand{\cov}{{\rm cov}}

\title{Analysis of Partially Observed Networks via Exponential-family Random Network Models}

\author[Ian E. Fellows]{Ian E. Fellows}
\email{ian.fellows@stat.ucla.edu}
\author[Ian E. Fellows and Mark S. Handcock]{Mark S. Handcock}
\address{Department of Statistics, University of California, Los Angeles, CA 90095-1554}
\email{handcock@stat.ucla.edu}

\begin{document}
\maketitle

\begin{abstract}
Exponential-family random network (ERN) models
specify a joint representation of both the dyads of a
network and nodal characteristics.  
This class of models allow the nodal characteristics to be modelled
as stochastic processes, expanding the range and realism of 
exponential-family approaches to network modelling. 
In this paper we develop a theory
of inference for ERN models when only part of the network is
observed, as well as specific methodology for missing data, including non-ignorable
mechanisms for network-based sampling designs and for latent class models.
In particular, we consider data collected via contact tracing, of considerable
importance to infectious disease epidemiology and public health.
\end{abstract}

\section{Introduction}

It is not uncommon for researchers to collect data on a subset of a single network rather than observing the full network. This partially observed case has been studied within the framework of exponential-family random graph models (ERGM) by \cite{hangile10aoas}, however their formulation suffers from the limitation that any nodal attributes included in the model must be fully observed, and only dyads may be missing. This assumption is not met in most sampling designs, where only some of the nodes are surveyed by the researcher, and reduces the practical usage of ERGMs in the missing data setting. 

By including nodal attributes as variates rather than fixed
quantities, exponential-family random network models
\citep[ERNM,][]{FellowsHandcock_2012} can
provide a convenient basis for inference in cases where the data
is partially unobserved, either due to design, or out-of-design (e.g., non-response) mechanisms.
While our framework is applicable to all partial observation mechanisms we consider three
common mechanisms for partial observations in more detail, specifically:

\begin{description}
\item[Missing Data:] If the population is comprised of a large number of units, or the number of edges is large, it is relatively common to find that the resources to observe a full network are not available. Often units or dyads are unavailable for sampling or do not provide complete responses to a survey instrument. In this case, only some of the dyads and nodal characteristics are collected.
We treat missing data as a form of sampling in which the sampling mechanism is unknown and
outside the control of the researcher, or an {\it out-of-design missing data mechanism}. 
A good example of this is the 
National Longitudinal Study of Adolescent Health (Add Health),
a school-based, longitudinal study of the health-related
behaviours of adolescents and their outcomes in young adulthood.
The study design
sampled 80 high schools and 52 middle schools from the U.S.,
representative with respect to region of country, urbanicity,
school size, school type, and ethnicity \citep{harris03}.
In 1994-95 an in-school questionnaire was administered to a
nationally representative sample of students in grades 7 through
12. In addition to demographic and contextual information,
each respondent was asked to nominate up to five boys and five girls
within the school whom they regarded as their best friends. Thus each student
could nominate up to ten students within the school \citep{udry03}. The nominations and contextual information were not available for some of the adolescents, either due to absence from school while the survey was being conducted, or refusal to participate. Thus, both the graph and nodal variates contained missing values.
\item[Network sampling designs:] Many studies in hard to reach populations use study designs that trace the linkages of an underlying social network. In these designs, the network is partially observed, however it is not of primary interest to the researcher. Such sampling designs have been exploited to estimate population disease rates \citep{gilehansocmeth10,gilejasa11,gilehan11MA}.
\item[Latent variables:] Some quantities of the network may be in principle unobservable. The probability model for a network may posit the existence of unknown variables which do not correspond to any observable quantity. For example, stochastic block models \citep{nowi:snij:2001} posit the existence of classes of nodes, conditional upon which the dyads are independent. These classes are unobserveable nodal characteristics and must be inferred from the relational data. Similarly, latent position cluster models \citep{handcock2006} posit the existence of unobservable continuous nodal quantities that provide  a spatial geometry for the network structure.
\end{description}

In this paper we develop approaches for each of these scenarios
in the context of ERNMs.  Sections \ref{sec:ernm} through
\ref{sec:mle} introduce ERNM and extend the theory to incorporate
partially observed populations.  Section \ref{se:specifics} develops methodology
for each of the scenarios. Sub-section \ref{se:monks} looks at the effect of
random non-response, and  sub-section \ref{ex:lcmonks}
applies a latent class model to extract unknown clusters from a
real data-set.  Sub-section \ref{ex:biasedseeds} develops estimates based
on contact tracing designs, which is of vital importance to the public health
community.  To our knowledge, the methods outlined in this paper
represent the first statistically
justifiable approach to inference from contract tracing data.

\section{Exponential-family random network models}\label{sec:ernm}

Exponential-family random network models \citep{FellowsHandcock_2012} are a generalisation of the exponential-family random graph model \citep{fra86,hunhan04}, where both dyads and nodal characteristics are treated as random variates. Formally, in a population of $n$ units, let $Y_{i,j}$ indicate that unit $i$ has a tie to unit $j$. Let $Y$ be an $n \times n$ matrix $[Y_{i,j}]$ and $X$ be a an $n \times K$ matrix $[X_{ik}]$ of unit covariates. We define a network $T$ as the union of the nodal covariates and the graph structure (i.e. $T = \{X,Y\}$).  An exponential family model of $T$ is expressed as

\begin{eqnarray}
P(T=t|\eta )  & = &  \frac{1}{c(\eta, {\cal T})}e^{\eta \cdot g(t)} ~~~~~ t\in {\cal T}, \label{eq:ernm}
\end{eqnarray}

where $\eta\in R^q$ is a vector of parameters, $g$ is a
$q-$vector valued function defining a set of sufficient
statistics, ${\cal T}$ is the sample space of networks and
$c(\eta, {\cal T}) = \sum_{t\in\cal T}{e^{\eta{\cdot}g(t)}}$ is
the normalising constant. This model is developed in \citet{FellowsHandcock_2012}.

\subsection{The Simple Homophily Model}\label{sec:simplehomo}

Though any set of network statistics can be represented by $g$ in equation \eqref{eq:ernm}, the examples in this paper will focus on a particularly parsimonious, but powerful, network model. Suppose that $X=(X_1,\ldots,X_n)$ is a univariate categorical variable with $m$ levels, labelled $0,\ldots ,m-1$. If $X_i = l$ we say that unit $i$ is in group $l$.  A joint model for $X$ and $Y$ is
$$
P(T=(y,x)|\eta )  =   \frac{1}{c(\eta, {\cal T})}e^{\eta_0\sum_{i,j}y_{i,j}  + \eta_2h(y,x)+ \sum_{l=0}^{m-2}\eta_{j+3}\sum_{i=1}^{n}I(x_i=l)}~~~~~(y,x) \in {\cal T}.
$$
The first term of this model is the number of edges, and controls the density of the graph. The last term represents the number of nodes in each category of $x$ ,except for the last level, which is dropped to maintain identifiability of the model. The second term $h$ is the regularised sample homophily of $x$, as introduced by \cite{FellowsHandcock_2012}, and is defined as
$$
h(y,x) = \sum_{k=0}^{m-1}\sum_{i:x_i=k} \sqrt{d_{i,k}(y,x)} - E_{\Perp}(\sqrt{d_{i,k}(y,x)}),
$$
where $d_{i,k}(y,x)$ is the number of edges between node $i$ and nodes in group $k$, and $E_{\Perp}(f(Y,X))$ is the expectation of the
statistic $f(Y,X)$, conditional upon $Y=y$ and the 
category counts (that is, the number of nodes in each category of $x$, $n(x) = \{n_k(x)\}_{k=1}^{K}$), assuming that $X$ and $Y$ are independent.
Thus, each term in the sum is the square root of the number of neighbours of a node which share the same category, minus what would be expected by chance. Using this form of homophily avoids the degeneracy problems found in other formulations. For a more thorough justification, see \citet{FellowsHandcock_2012}.

While the examples in this paper focus on applications of the simple homophily model, the framework presented here applies to any arbitrary set of network statistics $g$. For example, in many applications the nodal attributes are multivariate, and their relationships are of interest to the researcher. \citet{FellowsHandcock_2012} developed a network statistic that can be interpreted as a conditional logistic regression term which, if included, can model the relationship of several categorical variates.

\section{Likelihood-based Inference from Partially Observed Networks}

In this section we develop likelihood-based inference for network models based on partial observation of
the networks. The approach allows non-ignorable sampling mechanisms for the networks, including some common network-based sampling designs.
 
\cite{hangile10aoas} developed a theory of missing data for ERG models, and the specification for ERN models proceeds similarly, though our formulation supports a more general class of missingness processes known as missing not at random \citep[MNAR; see][]{rubin76}. Let $T_{obs}$ and $T_{miss}$ represent, respectively, the observed and unobserved part of the complete network $T.$. We write $T=(T_{obs}, T_{miss})$, with realisations $t=(t_{obs}, t_{miss})$. Let $W$ be a random variable representing the sampling process with realisation $w$. The probabilistic distribution of $W$ is the {\sl sampling mechanism}, and must fully specify the sample selection process, including the partition of $T$ into $T_{obs}$ and $T_{miss}$. Typically, $W$ will consist of an $n$ by $n$ matrix indicating whether the dyad was sampled, and an $n$ by $K$ matrix indicating which nodal attributes are missing; however, $W$ may contain additional information about the sampling, such as the order of sampling.

We write the {\sl full data likelihood} as

$$
p(T=t, W=w | \eta,\theta) =  p(W=w | T=t, \theta)\frac{1}{c(\eta, {\cal T})}e^{\eta \cdot g(t)},
$$
 and we wish to draw inferences about $\eta$ from the {\sl observed data likelihood}, defined as
\begin{equation}
p(T_{obs}=t_{obs}, W=w | \eta, \theta) = \sum_{t_{miss}} p(W=w | t=(t_{obs},t_{miss}), \theta)\frac{1}{c(\eta, {\cal T})} e^{\eta \cdot g((t_{obs},t_{miss}))}. \label{odlik}
\end{equation}

This probability model jointly represents the distribution of the network $T$, and the sampling process $W$. The functional form of $p(W=w | T=t, \theta)$ is dependent on the form of missingness, and will differ depending on how $T_{obs}$ was obtained. Section \ref{ex:biasedseeds} illustrates a design of particular interest known as biased seed link tracing. When the sampling probabilities only depend on the
observed data, then the sampling design is {\sl amenable} to the model \citep{hangile10aoas}, and is ignorable
in the sense of \citet{rubin76}. In this case, the likelihood simplifies to
\begin{eqnarray}
p(T_{obs}=t_{obs},W=w | \eta, \theta) &=& \sum_{t_{miss}} p(W=w | T_{obs}=t_{obs}, \theta)\frac{1}{c(\eta, {\cal T})} e^{\eta \cdot g((t_{obs},t_{miss}))}  \nonumber \\
 &=& p(W=w | T_{obs}=t_{obs}, \theta)\sum_{t_{miss}}\frac{1}{c(\eta, {\cal T})} e^{\eta \cdot g((t_{obs},t_{miss}))}  \nonumber \\
  &\propto& \sum_{t_{miss}} \frac{1}{c(\eta, {\cal T})} e^{\eta \cdot g((t_{obs},t_{miss}))}.
\end{eqnarray}

Thus, when the sampling process is ignorable, inferences on
$\eta$ are not affected by $p(W=w|T_{obs}=t_{obs},\theta)$, and
so knowledge of the sampling process is not essential for the process of inference.

Having defined the full and observed likelihood, it is also useful to define the {\sl missing data likelihood}:
$$
p(T_{miss}=t_{miss} | W=w, T_{obs}=t_{obs}, \eta,\theta) = \frac{ p(W=w | T=(t_{obs},t_{miss}), \theta)e^{\eta \cdot g((t_{obs},t_{miss}))}}{c(t_{obs}, w, \eta, \theta)}
$$
where
$$
c(t_{obs}, w, \eta, \theta) = \sum_{t_{miss}} p(W=w | T=(t_{obs},t_{miss}), \theta)e^{\eta \cdot g((t_{obs},t_{miss}))}.
$$

The (observed data) likelihood can then be rewritten as the ratio of two normalising constants
\begin{eqnarray}
p(T_{obs}=t_{obs},W=w | \eta, \theta) &=& \frac{1}{c(\eta, {\cal T})} \sum_{t_{miss}} p(W=w | T=(t_{obs},t_{miss}), \theta)e^{\eta \cdot g((t_{obs},t_{miss}))} \nonumber \\
&=&  \frac{c(t_{obs}, w, \eta, \theta)}{c(\eta, {\cal T})}, \nonumber 
\end{eqnarray}
and using this, we may write the observed data log likelihood ratio of $(\eta, \theta)$ versus $(\eta_0, \theta_0)$ as
\begin{eqnarray}
\ell(\eta,\theta) - \ell(\eta_0, \theta_0) &=& \log(\frac{c(t_{obs}, w, \eta, \theta)}{c(t_{obs}, w, \eta_0, \theta_0)}) - \log(\frac{c(\eta, {\cal T})}{c(\eta_0, {\cal T})}) \nonumber \\
&=& \log(\sum_{t_{miss}}\frac{p(W=w|T=t,\theta)}{p(W=w|T=t,\theta_0)}e^{(\eta-\eta_0)\cdot g(t)}\frac{p(W=w|T=t,\theta_0)e^{\eta_0\cdot g(t)}}{c(t_{obs}, w, \eta_0, \theta_0)}) \nonumber \\
& & - \log(\sum_{t_{miss}}e^{(\eta-\eta_0)\cdot g(t)}\frac{e^{\eta_0 \cdot g(t)}}{c(\eta, {\cal T})}) \nonumber \\
&=& \log(E_{\eta_0,\theta_0}(\frac{p(W=w|T,\theta)}{p(W=w|T,\theta_0)}e^{(\eta-\eta_0)\cdot g(T)}) | W=w, T_{obs}=t_{obs})\\ \nonumber
& & - \log(E_{\eta_0}(e^{(\eta-\eta_0)\cdot g(T)}))  \label{eq:lr}\\
&=& \log(E_{\eta_0,\theta_0}(e^{(\eta-\eta_0)\cdot g(T)}|T_{obs}=t_{obs}))-\log(E_{\eta_0}(e^{(\eta-\eta_0)\cdot g(T)})) \nonumber
\end{eqnarray}
\vskip -20pt
\begin{eqnarray}
&& + \log(\frac{E_{\eta,\theta~~}(P(W=w | T,\theta~) | T_{obs}=t_{obs})~~}{E_{\eta_0,\theta_0}(P(W=w | T,\theta_0) | T_{obs}=t_{obs})}).
\label{eq:lr2}
\end{eqnarray}

Both equation (4) 
and equation \eqref{eq:lr2} motivate
algorithms to draw inferences about $\eta$ and $\theta$.  Section
\ref{sec:mle} describes the algorithm motivated by equation (4), 
and Appendix A.1 
outlines an algorithm using equation \eqref{eq:lr2}.

\section{Calculating the MLE with MCMC} \label{sec:mle}

For most models, equation (4) 
is not analytically solvable. However we may approximate it by Markov Chain Monte Carlo (MCMC). Let $t^{(i)}$ and $t^{(i)}_m$ where $i \in (1,\dots ,M)$ be samples from the full likelihood and missing data likelihood respectively with parameters $\eta_0, \theta_0$. Then equation (4) 
may be approximated by

\begin{equation} \label{eq:lra}
\ell(\eta,\theta) - \ell(\eta_0, \theta_0) \approx \log(\frac{1}{M}\sum_i^M\frac{p(w|t^{(i)}_m,\theta)}{p(w|t^{(i)}_m,\theta_0)}e^{(\eta-\eta_0)\cdot g(t^{(i)}_m)}) - \log(\frac{1}{M}\sum_i^Me^{(\eta-\eta_0)\cdot g(t^{(i)})})
\end{equation}

As $\eta,\theta$  move away from $\eta_0,\theta_0$ the quality of this approximation degrades. Because we will be optimising equation (4), 
it is useful to have both the first and second derivatives of the log likelihood, which are

$$
\frac{\delta \ell}{\delta \eta} = E_{\eta,\theta}(g_i(t)| T_{obs}=t_{obs},W=w) - E_{\eta,\theta}(g_i(T))
$$
$$
\frac{\delta^2 \ell}{\delta \eta_i \delta \eta_j} = -\cov(g_i(T),g_j(T))  + \cov(g_i(T),g_j(T) | T_{obs}=t_{obs},W=w).
$$
The expectations and covariances in these derivatives can be approximated using the conditional and unconditional MCMC samples and thus we can then use the following algorithm to approximate the MLE.

\begin{enumerate}
\item Let $k=0$ and choose initial parameter values $\eta^{(0)}$, $\theta_0$.
\item Use MCMC to generate k samples, $t^{(i)}_{miss}$ from $P(T_{miss}=t_{miss}|\eta^k, T_{obs}=t_{obs}, W=w)$.
\item Use MCMC to generate m samples $t^{(i)}$ from $P(T=t | \eta^k )$.
\item Using the samples from step 2 and 3 in equation \eqref{eq:lra}, find $\eta^{k+1}, \theta^{k+1}$ maximising the likelihood ratio, subject to $||\eta^{k+1} - \eta^k|| < \epsilon$ and $||\theta^{k+1} - \theta^k|| < \epsilon$.
\item If the likelihood has not converged, set $k=k+1$ and go to step 2.
\item Let the MLE estimate be $\hat{\eta} = \eta^{k+1}$ and $\hat{\theta} = \theta^{k+1}$ 
\end{enumerate}

Asymptotic standard errors for $\hat{\eta}$ may be obtained using an MCMC approximation to the Fisher information (i.e. the second derivative of the log likelihood). While asymptotics of the Fisher information are not assured with respect to ERNM (or ERGM) models, \cite{FellowsHandcock_2012} show strong empirical agreement between the Fisher information standard errors and parametric bootstrap simulations. Standard errors for the mean value parameters $\hat{\mu} = E(g(T)|\eta=\hat{\eta})$ can be approximated by MCMC sampling.

\section{Specific forms of partial observation}\label{se:specifics}
In this section we consider the three common forms of partial observation
considered in the introduction, each corresponding to a different mechanism of partial observation
or conceptualisation of that mechanism.

\subsection{Missing Data: Unobserved Relational Information}\label{se:monks}

\begin{figure}
  \noindent 
  \begin{flushleft}
    \includegraphics[scale=.32]{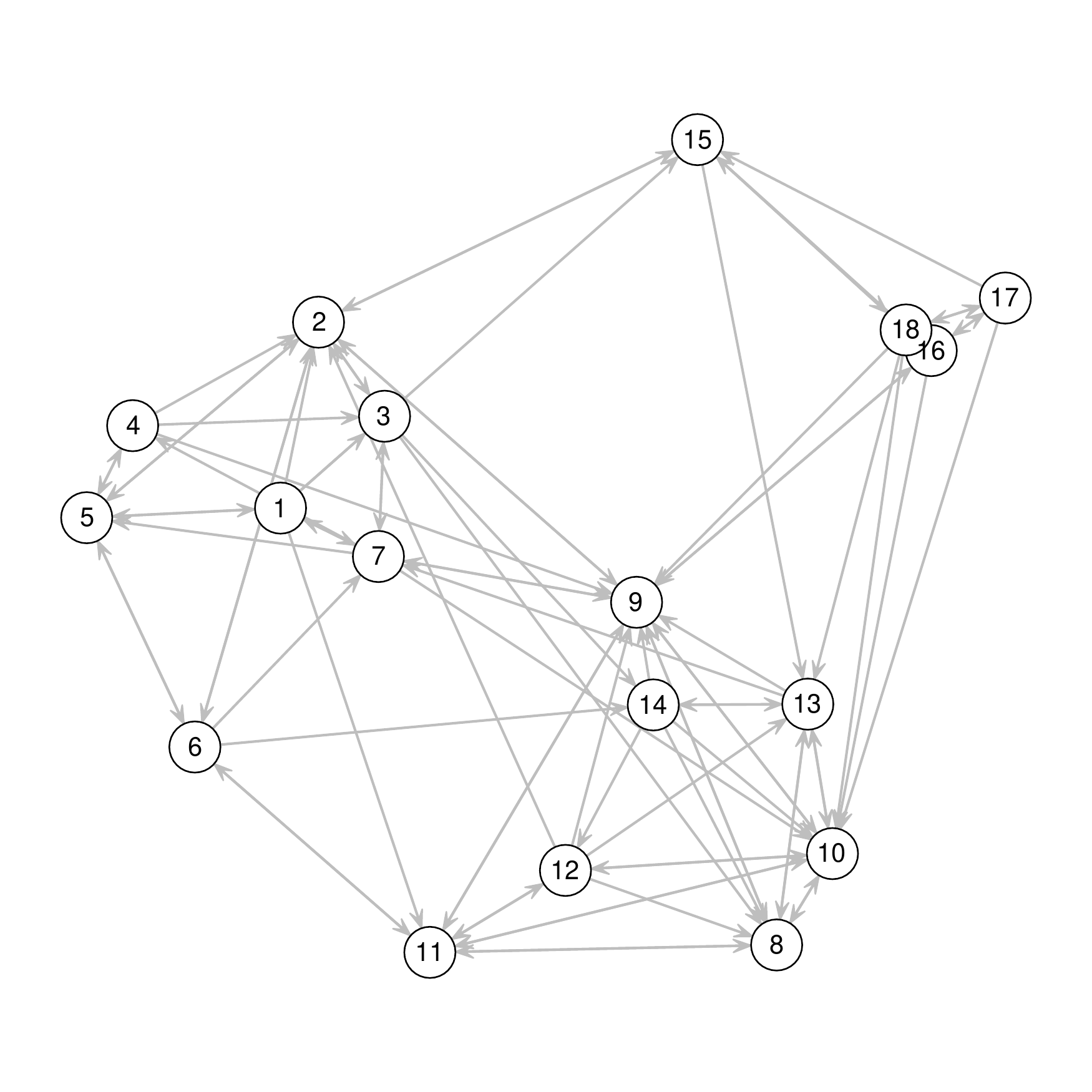}
  \end{flushleft}
     \noindent 
      \begin{flushright}
        {\scriptsize \vskip -17em}{\small
          \begin{tabular}{rlrl}
            1 & Ramauld (L) & 10 & Gregory (T) \\
            2 & Bonaventure (L) & 11 & Hugh (T) \\
            3 & Ambrose (L) & 12 & Boniface (T) \\
            4 & Berthold (L) & 13 & Mark (T) \\
            5 & Peter (L) & 14 & Albert (T) \\
            6 & Louis (L) & 15 & Amand (O) \\
            7 & Victor (L) & 16 & Basil (O) \\
            8 & Winfred (T) & 17 & Elias (O) \\
            9 & John (T) & 18 & Simplicius (O) \\
          \end{tabular}
        } 
      \end{flushright}
      \caption{\label{fig:Monks-dataset}Relationships among monks
        within a monastery and their affiliations as identified by
        Sampson: Young (T)urks, (L)oyal Opposition, and (O)utcasts.}
\end{figure}

It is common when surveying networked populations that there are insufficient resources to conduct a
census of the population and their relations. For efficiency reasons, a sampling based survey is
undertaken, or the full network is partially observed due to non-response. 
In this sub-section, we give an illustration of the effect of non-response where the dyad information is
missing completely at random.
We consider the relations of ``liking'' among 18 monks in a monastery
\citep{sampson69}. The network analysed has a directed edge between
two monks if the sender monk ranked the receiver monk in the top three
monks for positive affection in any of the three interviews given over
a twelve month period \citep{hrh02}.  The sociogram of this data-set is shown in
Figure \ref{fig:Monks-dataset}.
One nodal attribute of interest is an indicator of attendance at the minor ``Cloisterville" seminary before coming to the monastery.

\begin{figure}
\vspace*{.1in}
\begin{center}
\includegraphics[scale=.2]{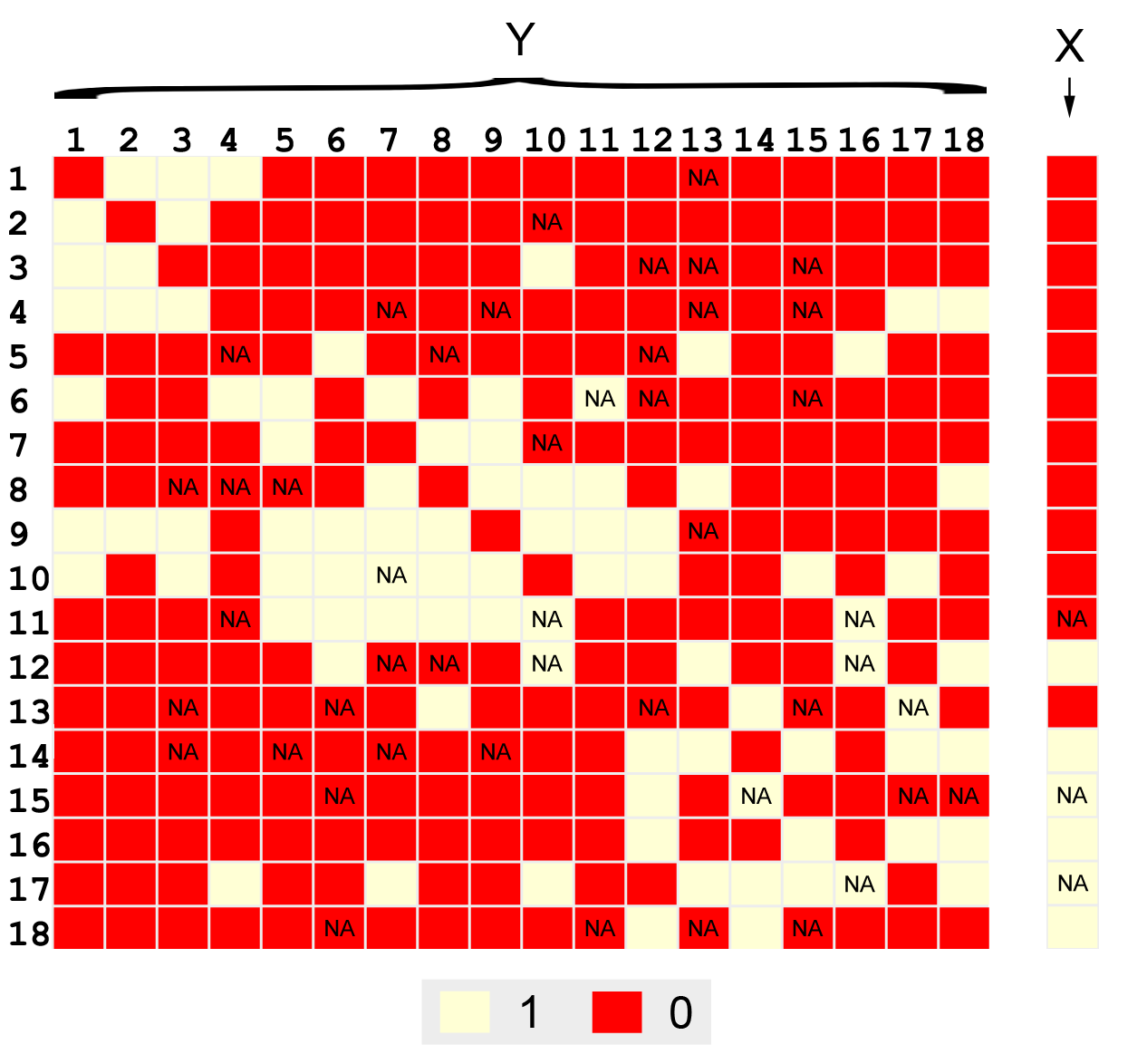}
\end{center}
\caption{
Sampson's monk's with 15\% missingness. Cloisterville status marked on the right hand side.}\label{fig:mcar1}
\end{figure}

\begin{figure}
\vspace*{.1in}
\begin{center}
\includegraphics[scale=0.7]{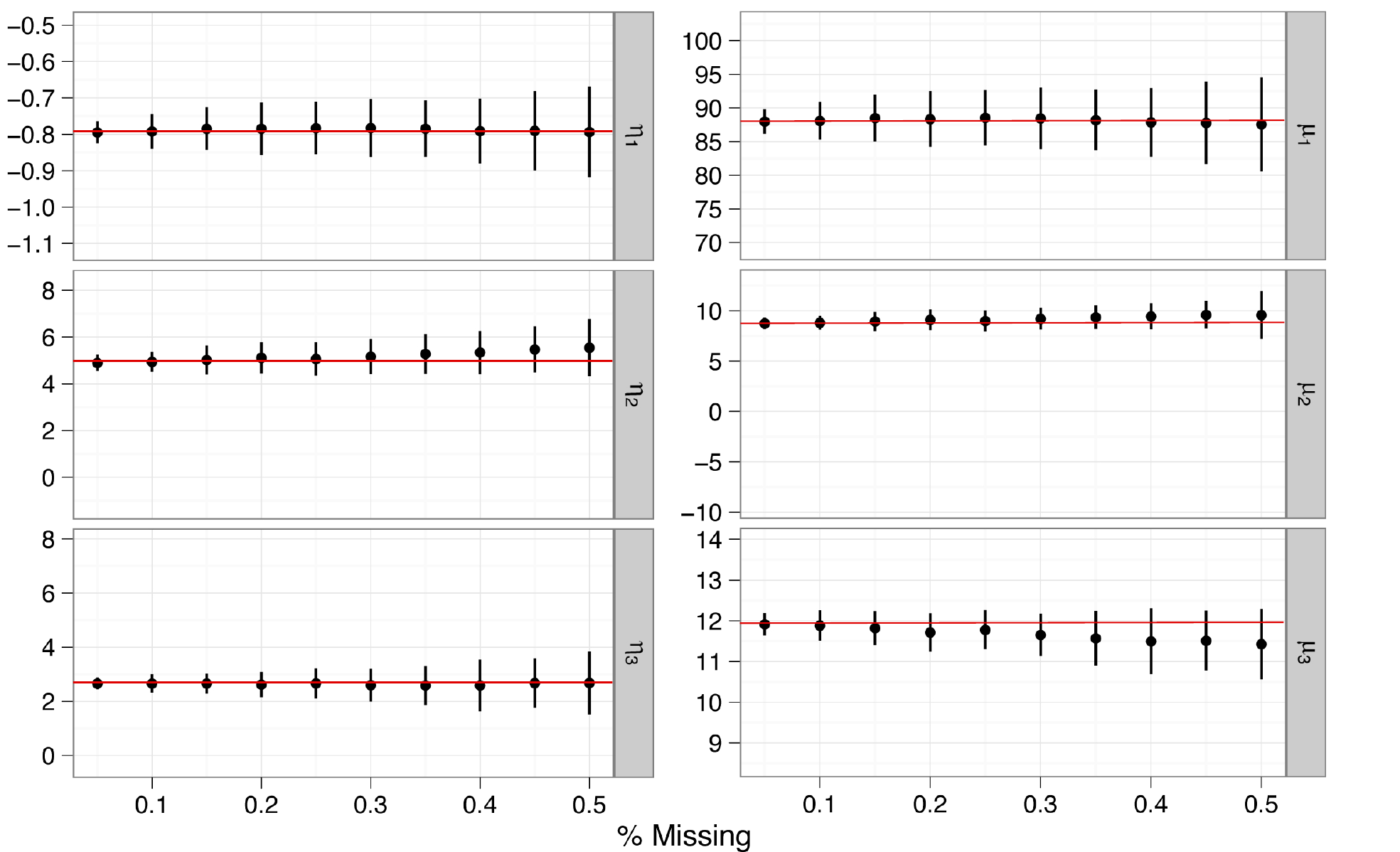}
\end{center}
\caption{
Means and standard deviations of model estimates. Red lines indicate fully observed MLE}\label{fig:mcar2}
\end{figure}

We fit a simple homophily model on Cloisterville status using the full data. We then ran simulations on the effect of missingness by selecting dyads, and Cloisterville status variates, completely at random and setting them to missing. Figure \ref{fig:mcar1} shows one simulated missingness pattern with 15\% missing. We ran 100 simulations at each missingness percentage. Means and standard deviations of the ERNM models fit to these simulated missingness patterns are displayed in Figure \ref{fig:mcar2}. 

We see that the standard deviations of the estimates increase as the amount of missingness increases. At the higher missingness levels some bias is apparent relative to the full data MLE, but not more than one standard deviation. One possible explanation for this bias is that there were only six monks who attended Cloisterville, and so at 50\% missingness, a significant number of samples will include no (or perhaps a single) Cloisterville monks. 

\subsection{Latent Variables: Stochastic Block Models}\label{ex:lcmonks}

In this sub-section we consider the situation where some characteristics of the network are posited 
but unobserved. Specifically, we consider the case where each node of the network 
belongs to a latent class, and the structure of the network depends on that latent class.
The traditional approach to this has been stochastic block models \citet{nowi:snij:2001}, and here
we show how these models fall naturally out of our general formulation.

It is apparent from Figure \ref{fig:Monks-dataset} that the pattern of ``liking" between the
monks may exhibit clustering. 
Through close sociological study, \cite{sampson69} identified three clusters
which he dubbed the Turks, Loyal Opposition and the Outcasts
(see: Figure \ref{fig:Monks-dataset}).  Here we will attempt to
identify clusters by inferring class membership from the graph.  We fit
the simple homophily model of Section \ref{sec:simplehomo} to this data, assuming a class
covariate, $X$, with three levels, and that all of the monks are
``missing" their class covariate.  The simple homophily model
treated this way represents a novel latent block model in the
spirit of \cite{nowi:snij:2001}.  Note that the missingness
process here is ignorable because it does not depend on
unobserved quantities as all of the $x$ values are missing
regardless of the $Y$ values. We fit the model using the algorithm in
Section \ref{sec:mle}.  Table \ref{tbl:samp1} shows the
maximum likelihood parameter estimates, along with standard errors of the estimators based on the Fisher information.

\begin{table}
\caption{Latent Class model for Sampson's monks. \label{tbl:samp1}}
\centering
\begin{tabular}{lrrrr}
  \hline
Term & $\hat\eta$ & $\hat\mu$ & ${\rm s.e.}(\hat\eta)$ & ${\rm s.e.}(\hat\mu)$\\ 
  \hline
  \# of edges &  -0.58 &  88.23 & 0.14 & 7.48\\ 
  Homophily & 7.28 & 15.30  & 0.91 & 1.33\\ 
  \# in group 0 & -2.50 & 3.95  & 1.44 & 1.08\\ 
  \# in group 1 & -0.02 & 6.95 & 1.31 & 0.99\\ 
  \hline
\end{tabular}

\end{table}

The natural parameter estimates indicate significant homophily in tie formation
based on the class. It also indicates that the number of monks in the third class is
significantly more than those of the other two classes, which are not statistically significantly
different in size. The mean value parameters indicate that the expected number of ties is about
88, and the expected numbers in the three groups are 4, 7 and 7.

An advantage of this approach is that we can investigate the probability of class membership, which
is well defined through our framework as $p(X=x|Y=y_{obs},\eta)$.
To compute $p(X=x|Y=y_{obs},\eta)$ we simulated a large number of samples from
$p(X=x|Y=y_{obs},\hat{\eta})$ using MCMC to show the probability of the monks being in the 
classes displayed in Figure \ref{fig:Monks-dataset} to be above 0.9999.
These clusters were also identical to those chosen by \citet{sampson69} and 
verified by later research \citet{brei:boor:arab:1975, handcock2006}. 

In addition to assuming a set number of latent classes for the
model, we can also use the MLE procedure to select an appropriate
number of clusters for the data.  We fit the simple homophily
model with a latent variable $X$ able to take a potentially large
number of values (e.g., the number of monks).  In this case
$p(X=x|Y=y_{obs},\hat{\eta})$ places zero mass for all but three of the
groups.  This is evidence that the three groups we have
identified are a good classification for these data.  More
sophisticated model selection approaches for choosing the number
of clusters are possible \citep{handcock2006}, and are left for future work.

Our form of the stochastic block model is conceptually very clean with the ability to naturally
incorporate additional covariates, multiple membership variables,
and extensions to an unbounded numbers of classes. Inference
is straightforward, and quantities such as the probability of class membership are well defined and
interpretable. We leave a full exploration of these for latter work.

\subsection{Network Sampling: Biased Seed Link-Tracing}\label{ex:biasedseeds}
Handcock and Gile (2010) explored the idea of sampling networks by tracing the edges. As a general concept, link tracing involves selecting one or more seed nodes, and then observing the edges connected to those seeds. One or more of these edges are then followed to the neighbouring node, whose ties are observed, and the process is continued. Each iteration of this process is known as a wave.

Provided that the seed nodes are chosen at random, and the method by which edges are chosen to be followed depends only on the observed data, this missingness process is ignorable. To be explicit, consider a link tracing process with $k$ waves. Let $w_i$ be the ordered set of nodes and edges sampled in the $i$th wave in the order in which they were sampled, $w = \{w_0,...,w_k\}$, and $w_{-i}=\{w_0,...,w_{i-1},w_{i+1},...,w_k\}$. If the seeds are chosen at random, and the edges followed by the sampling process are also chosen at random, then $p(W=w | T=t,\theta) = p(W=w | T_{obs}=t_{obs},\theta)$, implying that the missingness is ignorable.

In many cases, however, the seeds are not chosen at random from the population, but are some form of convenience sample. For example, in a population where some people have an infection and others do not, we may start with a sample of $s_i$ seeds picked at random from among the infected individuals, and $s_{-i}$ seeds picked from the non-infected individuals. These seeds are then used as a starting point for standard link tracing. We may then write the sampling probability as
\begin{eqnarray}
p(w | t,\theta) &=& p(w_0 | t,\theta)p(w_{-0} | t_{obs},w_0,\theta) \nonumber \\
&=& \frac{(n_i-s_i)!}{n_i!}\frac{(n_{-i}-s_{-i})!}{n_{-i}!}p(w_{-0} | t_{obs},w_0,\theta), \nonumber 
\end{eqnarray}
where $n_i$ and $n_{-i}$ are the number of infected and non-infected in the population, respectively. Note that $p(w_{-0} | t_{obs},w_0,\theta)$ does not depend on $t_{miss}$ and may be factored out of the likelihood in equation \eqref{odlik}. Thus there is no need to calculate $p(w_{-0} | t_{obs},w_0,\theta)$ explicitly, as it makes no impact on the likelihood.
Hence, in this case, we can compute the likelihood without knowing the specific mechanism of seed
selection.

\subsection{Network Sampling: Positive Contact Tracing}\label{ex:contacttracing}

As emerging epidemics develop, control measures (e.g., treatment, isolation and culling) 
focus on those members of the population that are known to have the infection. Because there
are often many infected people who are unobserved, control can be ineffective
(e.g., HIV \citep{potterat1989}.
The alternative of applying control measures to the entire population can be economically
infeasible or ineffective (e.g., some instances of safe sex education)
\citep{potterat1989,klinkenberg2006}.
Contact tracing is the hybrid approach of treating both the known infected individuals and those who may have been infected by them \citep{potterat1989,klinkenberg2006}. 
In U.S. public health, health clinics are
required by state law to notify those at risk from infection due to their sexual
relations with individuals tested, and found to be infected, by the clinic.
The process of locating, notifying and then testing
partners that may have been exposed to an infectious agent allows additional information about the partners
to be collected. 
While the primary purpose of contact tracing is
disease control via partner notification and partner services, it is also a form of data
collection that is rarely utilised.
Such approaches are used most commonly for syphilis and
HIV/AIDS, but also for other STIs such as gonorrhea and chlamydia \citep{golden2004}, as well as
 routinely for tuberculosis and infectious disease outbreaks.
Contact tracing has also been applied in many recent epidemics
\citep{fenner1988,ferguson2001,Donnelly2003}. 
In {\sl positive contact tracing}, we follow all edges from infected nodes, but edges from uninfected nodes are not followed.

While the process varies from state to state and also by disease, we consider
the following biased seed link tracing process:

\begin{enumerate}
\item Select $s_{-i}$ seed subjects at random from among the non-infected population, observe
them.
\item Select $s_i$ seeds subjects at random from among the infected population, observe them.
\item Choose the next infected seed at random.
\item Observe all edges from the selected subject, and the infection status of these subjects.
\item For all infected neighbours of the selected subject, go to step 4.
\item If all the seeds have not been chain sampled, go to step 3
\end{enumerate}

\begin{figure}
\vspace*{.1in}
\begin{center}
\includegraphics[scale=0.5]{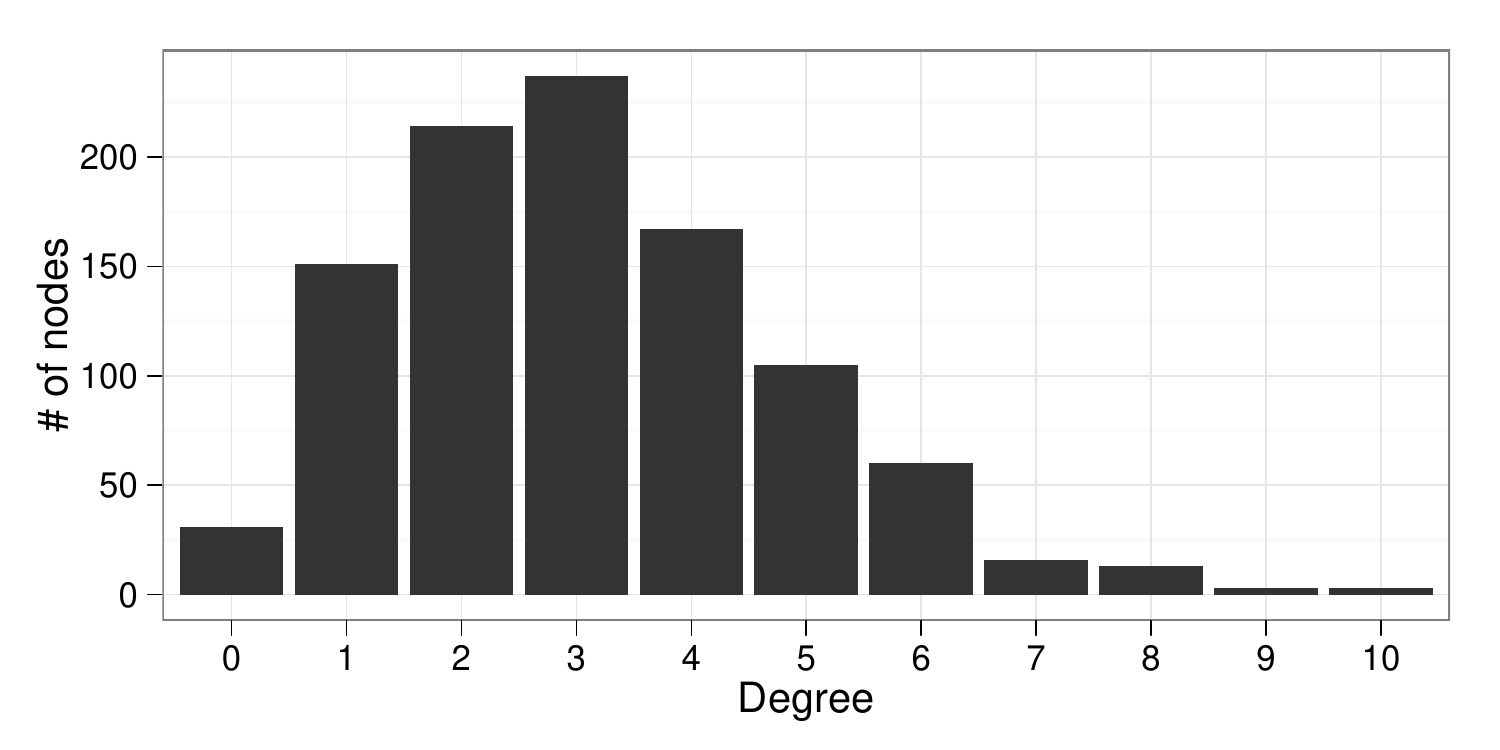}
\end{center}
\caption{
Degree distribution of the networked population.}\label{figure:f1}
\end{figure}

\begin{figure}
\vspace*{.1in}
\begin{center}
\includegraphics[scale=.5]{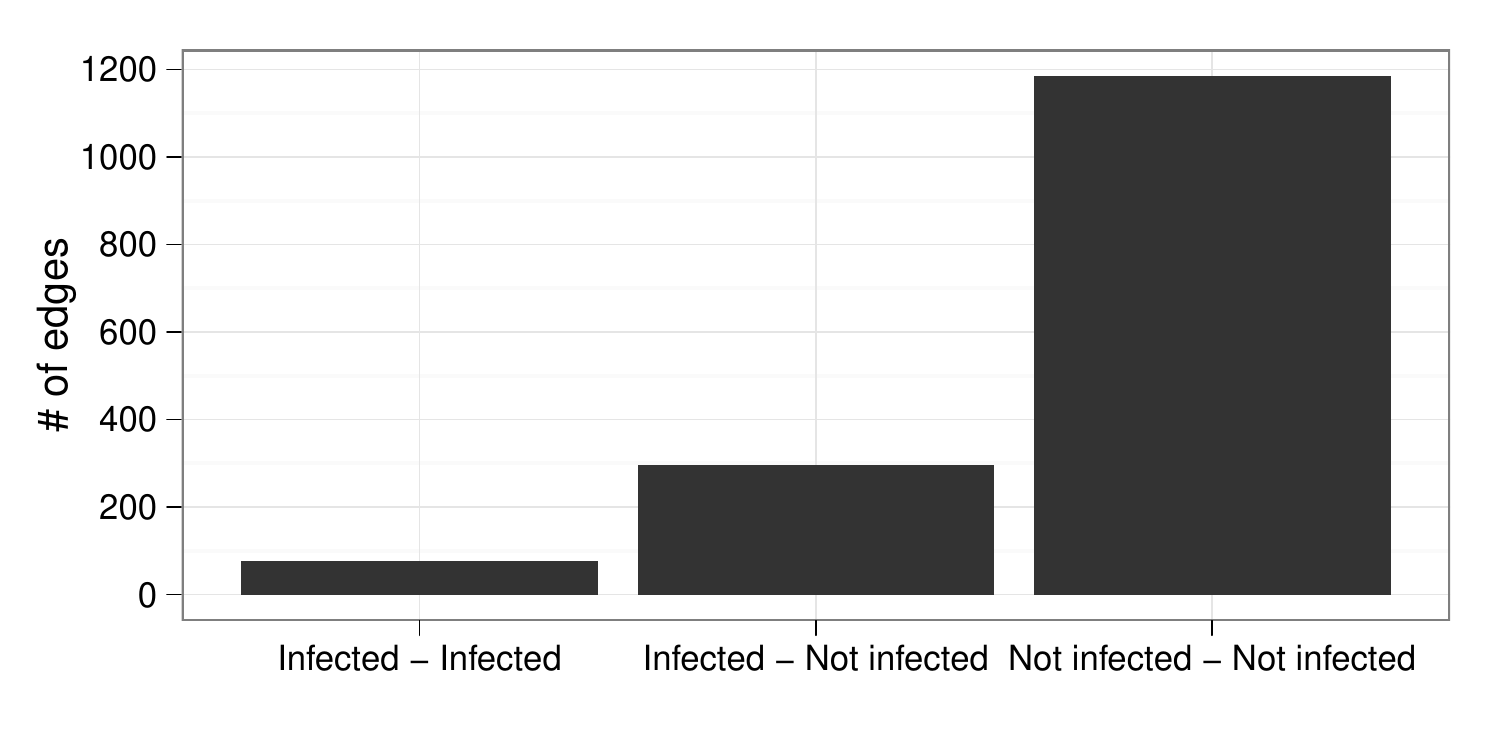}
\end{center}
\caption{
Mixing statistics: Counts of the numbers of edges by the infection status of the incident nodes for the networked population.}\label{figure:f2}
\end{figure}

We simulated a networked population of $n=1000$ people from the simple homophily
model of Section \ref{sec:simplehomo} with natural parameters of $\eta = (-5.8, .7, -1.95)$. 
The number of infected nodes was fixed at 150.
The generated network had a mean degree of 3.1, and its degree distribution is displayed in Figure \ref{figure:f1}. There were 296 infected to non-infected ties, with the mixing distribution displayed in Figure \ref{figure:f2} indicating moderate homophily.

\begin{figure}
\vspace*{.1in}
\begin{center}
\includegraphics[scale=.7]{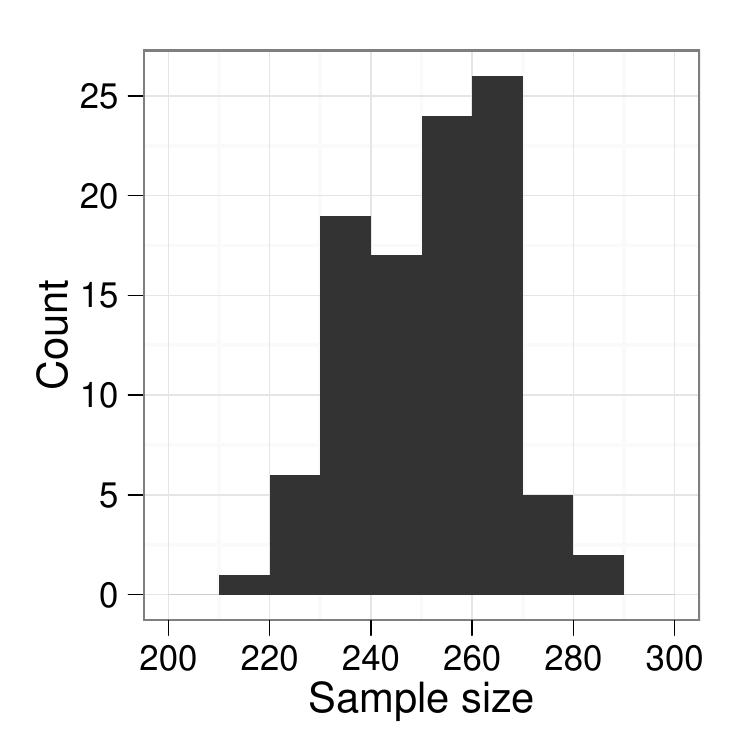}
\end{center}
\caption{
Sizes of the contact-traced samples based on 40 seed subjects $(s_i=40, s_{-i}=0)$.}\label{figure:f3}
\end{figure}

Starting with $s_i=40$ infected seeds, we simulated 100 positive link tracing
samples for each of $s_{-i}=(0,45,90,135,180,225)$.  Figure
\ref{figure:f3} displays a histogram of the sizes of the samples
when there are no non-infected seeds (i.e., $s_{-i}=0$).

To provide a comparison for our method we considered two estimators that could be utilised. Neither of
them uses a model for the networked population but is motivated by approximations to the sampling
design. The first treats the sample as a simple random sample
$$
{\rm Naive} = n\frac{n_i}{n_i+n_u},
$$
where $n_i$ and $n_u$ are the number of infected and uninfected in
the sample respectively.
The second adjusts for the sampling of the seeds
$$
{\rm Naive\ (seed\ adj.)} = (n-s_i-s_{-i})\frac{n_i-s_i}{n_i-s_i+n_u-s_{-i}}+s_i.
$$
Our approach is to fit an ERNM to the contact tracing data. In this situation the contact tracing
sampling design is clearly informative. For comparison, 
we compute two estimates of the model. The first takes into account
the informativeness of the contact tracing design (MNAR) and the other
assume it is ignorable (MAR).  
These are based on the likelihoods 
\ref{odlik} and \ref{eq:lr}, respectively, and the algorithm in Section \ref{sec:mle}.

\begin{figure}
\vspace*{.1in}
\begin{center}
\includegraphics[scale=.9]{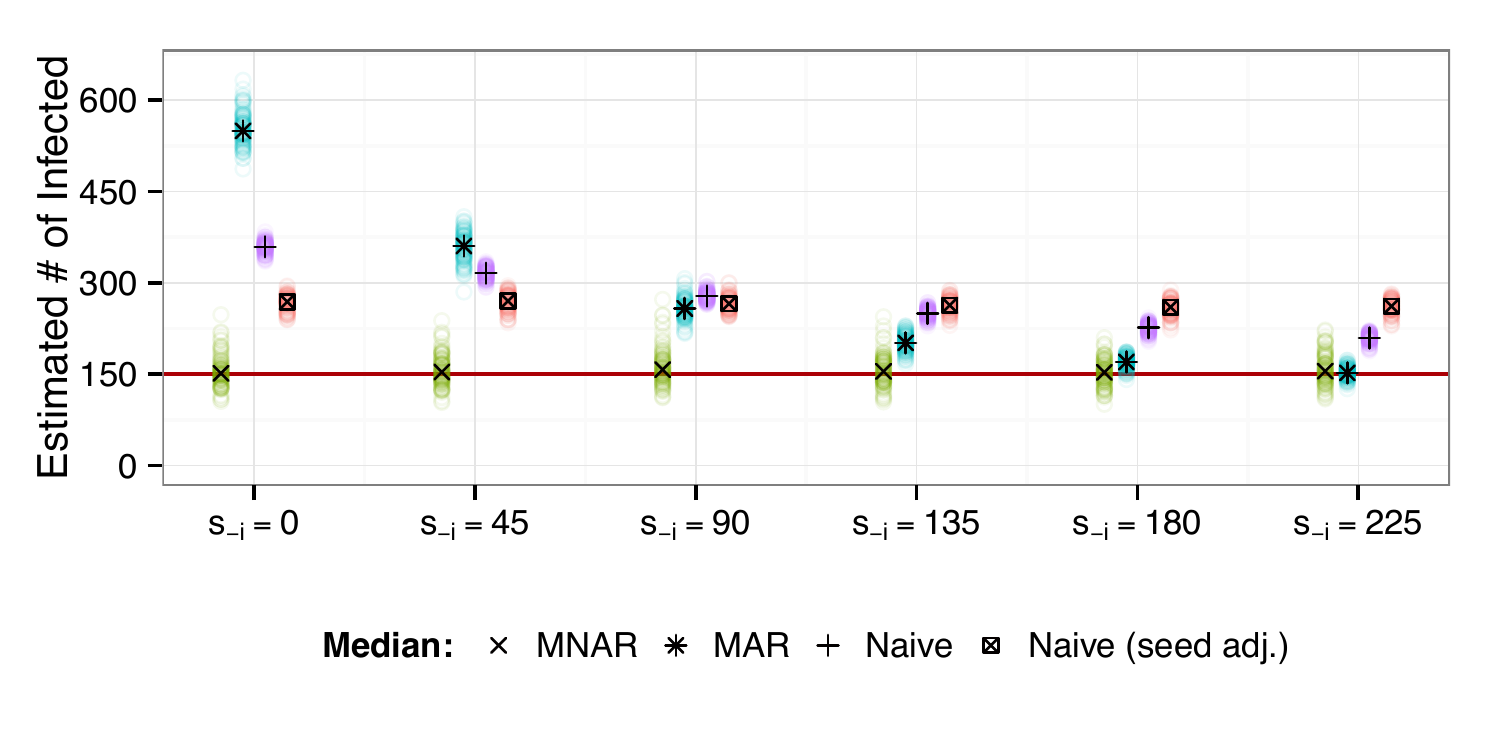}
\end{center}
\caption{
Estimates via contact tracing with $s_i=40$ infected seeds and varying numbers of non-infected
seeds.}\label{figure:f4}
\end{figure}

Figure \ref{figure:f4} shows the results for each of the estimators over the
samples.  The median of the MNAR estimator is
centred around the true value of 150 in all sampling scenarios,
while the MAR estimator performs poorly with all infected seeds ($s_{-i}=0$) and
increasingly well as the number of non-infected seeds increases to $s_{-i}=225$.
This is somewhat
expected as the proportion of infected in the seeds approximately
matches that of the population when $s_{-i}=225$.  The two naive
estimators are significantly biased across all samples. This is especially true for the sample mean
which is biased both by the seed selection and by the link-tracing design. The adjusted sample mean
corrects somewhat for the seed bias but does not represent the link-tracing. 

This application illustrates the advantage of the model-based approach
over the {\sl ad hoc} estimators. By representing the structure of the networked population, the
model-based approach can leverage the information in the data more efficiently.

\section{Discussion}

In this paper we have given a concise and systematic statistical
framework for dealing with
partially observed network data when some knowledge is available on the 
sampling design. The framework
includes, but is not restricted to, ignorable sampling designs.
We have also shown that likelihood-based inference is practical under partial
observation for ERN models, and that the likelihood framework naturally accommodates standard
sampling designs.

We developed and implemented algorithms to compute Monte Carlo approximations to the
 likelihood, and showed how these can be used in practice. Three important special cases of these designs were demonstrated in Section \ref{se:specifics}. 
In Sub-section \ref{se:monks} we consider a missingness process which randomly selected dyads and nodal attributes to be missing. Sub-section \ref{se:monks} considers the case where {\it all} nodal attributes are missing, thus introducing a novel form of the latent cluster model. 

In Sub-section \ref{ex:biasedseeds} we consider
non-ignorable sampling in the context of contact tracing data, a case of vital importance to public health. At present, this is the first statistically defensible approach to inference in this form of data. The example presented here shows that the MLE estimation task is robust, in that it can be applied successfully to moderately large networks (1000 nodes), with significant missingness (>70\% of nodes unobserved), but is limited by the fact that inference was performed on a simulated network. Whether the model presented here would provide a good fit for real public health data remains an important research question that we hope to address in the future.

\appendix
\section*{Appendix: Algorithmic and Computational Details} \label{sec:append}
\subsection*{A.1:~~~ Alternate MLE Formulation } \label{subsec:altmle}

While the algorithm outlined in Section \ref{sec:mle} works well, there are some situations where an alternate formulation using equation \eqref{eq:lr2} may be useful. First let us consider the case where $\theta=\theta_0$, then the likelihood is

\begin{equation}\label{eq:aplr}
\ell(\eta) - \ell(\eta_0) =\log(E_{\eta_0}(e^{(\eta-\eta_0)\cdot g(T)}|t_{obs}))-\log(E_{\eta_0}(e^{(\eta-\eta_0)\cdot g(T)})) +
\log(\frac{E_{\eta}(P(W=w | T,\theta) | T_{obs} = t_{obs})}{E_{\eta_0}(P(W=w | T,\theta) | T_{obs} = t_{obs})} 
\end{equation}

The first expectation, and the expectation in the denominator of the third term, can be calculated using an MCMC sample from $p(t | t_{obs},\eta_0)$. The second can be approximated with an MCMC sample from $p(t|\eta_0)$. The numerator of the third term can be approximated by importance sampling.

$$
E_{\eta}(P(W=w | T,\theta) | T_{obs} = t_{obs}) \approx \frac{1}{k}\sum_i^k p(w | t^{(i)},\theta)\omega^{(i)}
$$

where $t^{(i)} \sim p(t | t_{obs},\eta_0) $ and 
$$
\omega^{(i)} = \frac{e^{(\eta-\eta_0)\cdot g(t^{(i)})}}{\sum_j^k e^{(\eta-\eta_0)\cdot g(t^{(j)})}}
$$
If the sampling process is ignorable, then the third term drops out of the likelihood ratio. The first and second derivatives of the likelihood are useful in the maximisation process. For notational convenience let $\Delta_i(t) = g_i(t)-E(g_i(T))$.

\begin{eqnarray}
\frac{\delta \ell}{\delta \eta}&=&\frac{\delta}{\delta \eta_i} \log(\sum_{t_{miss}}\ p(W=w|T=t)P(T_{miss}=t_{miss}|\eta, T_{obs}t_{obs})P(T_{obs}=t_{obs}|\eta))\nonumber \\
&=&\frac{\sum_{t_{miss}}\ p(W=w|T=t)\Delta_i(t)P(T_{miss}=t_{miss}|\eta, T_{obs}=t_{obs})P(T_{obs}=t_{obs}|\eta)}{\sum_{t_{miss}}\ p(W=w|T=t)P(T_{miss}=t_{miss}|\eta, T_{obs}=t_{obs})P(T_{obs}=t_{obs}|\eta)}\nonumber \\
&=& \frac{ E(p(W=w|T)\Delta_i(T)| T_{obs}=t_{obs}) }{ E(p(W=w|T)|T_{obs}=t_{obs}) } \nonumber
\end{eqnarray}

\begin{eqnarray}
\frac{\delta^2 \ell}{\delta \eta_i \delta \eta_j}&=&\frac{\delta}{\delta \eta_j} \frac{
\sum_{t_{miss}}\ P(W=w | T=t)\Delta_i(t)P(T_{miss}=t_{miss}|\eta, T_{obs}=t_{obs})P(T_{obs}=t_{obs}|\eta)
}{
\sum_{t_{miss}} P(W=w | T=t)P(T_{miss}=t_{miss}|\eta, T_{obs}=t_{obs})P(T_{obs}=t_{obs}|\eta)} \nonumber \\
&=& -\cov(g_i(T),h_j(T)) + \frac{E(p(W=w|T)\Delta_i(T)\Delta_j(T)|T_{obs}=t_{obs})}{E(p(W=w|T)|T_{obs}=t_{obs})} \nonumber \\
&& - \frac{E(p(W=w|T)\Delta_i(T) | T_{obs}=t_{obs})E(p(W=w|T)\Delta_j(T) | T_{obs}=t_{obs})}{E(p(W=w|T)| T_{obs}=t_{obs})^2} \nonumber
\end{eqnarray}

And if the missingness process is ignorable, these equations simplify to
$$
\frac{\delta \ell}{\delta \eta} = E(\Delta_i(T)| T_{obs}=t_{obs})
$$
$$
\frac{\delta^2 \ell}{\delta \eta_i \delta \eta_j} = -\cov(g_i(T),g_j(T))  + \cov(g_i(T),g_j(T) | T_{obs}=t_{obs}) 
$$

If we fix $\eta$, then the observed likelihood of $\theta$
\begin{eqnarray}
L(\theta | t_{obs} , w,\eta) &\propto& P(t_{obs}|\eta) E(P(W=w | T,\theta) | T_{obs} = t_{obs})\nonumber \\
&=&E(P(W=w | T,\theta) | T_{obs} = t_{obs},\eta) \nonumber
\end{eqnarray}
can be maximised to find the MLE of $\theta$.

This motivates the following algorithm for maximising the observed data likelihood.

\begin{enumerate}
\item Let $k=0$ and choose initial parameter values $\eta^{(0)}$, $\theta_0$.
\item Use MCMC to generate k samples, $t^{(i)}_{miss}$ from $P(t_{miss}|\eta^k, t_{obs})$.
\item Use MCMC to generate m samples $t^{(i)}$ from $P(t | \eta^k )$.
\item Set $\theta^{k+1} = \argmax(E(P(w | T,\theta) | T_{obs} = t_{obs},\eta))$, with samples from step 2 used to approximate the expectation.
\item Using the samples from steps 2 and 3 to approximate the relevant expectations, find $\eta^{k+1}$ maximising equation \eqref{eq:aplr} subject to $||\eta^{k+1} - \eta^k|| < \epsilon$.
\item Set $k=k+1$, and go to step 2.
\end{enumerate}

The disadvantage of this method is that if the networks generated by the MNAR process are very different from those generated assuming MAR, the estimates of the last expectation in equation \eqref{eq:aplr} can become unstable. The benefit of using this method is that the sampling probability ($P(W=w | T=t,\theta)$) only needs to be calculated for networks included in the sample, and not at every MCMC step as is required by the algorithm in Section  \ref{sec:mle}, so if the sampling probability is computationally expensive to calculate, this method can be significantly faster than the one outlined in Section \ref{sec:mle}

\subsection*{A.2:~~~ Estimating Network Statistics}

We can use MCMC samples from $p(t_{miss} | t_{obs}, \eta)$  to estimate the network statistics of the sampled network. Suppose that we have used MCMC to draw $k$ samples $t_{miss}^{(i)}$ from the distribution $p(t_{miss} | t_{obs}, \eta)$, and $t^{(i)} = (t_{obs}, t_{miss}^{(i)})$. Then we can estimate the expectation of a set of network statistics $g$ as
$$
E(g(T)|t_{obs},\eta) \approx \frac{1}{k}\sum_{i=0}^kg(t^{(i)}).
$$
However, this equation ignores the possible bias introduced by our sampling process $w$. The distribution that we should be sampling from is the full conditional distribution of $t_{miss}$,

$$
p(T_{miss}=t_{miss} | T_{ob}=t_{obs},W=w, \eta) \propto p(T_{miss}=t_{miss} |T_{obs}= t_{obs}, \eta)p(W=w|T=t,\theta).
$$

We then use importance sampling to estimate the relevant quantity

$$
E(g(T)|t_{obs},w,\eta,\theta) \approx \frac{\sum_{i=0}^kg(t^{(i)})p(W=w|T=t^{(i)},\theta)}{\sum_{i=0}^kp(W=w|T=t^{(i)},\theta)}.
$$

\bibliographystyle{jrss}
\bibliography{mnar}

\end{document}